\documentclass{iau_FM}
\usepackage{graphicx}

\title[Magnetism in the SKA Era] 
{Magnetism in the \\ Square Kilometre Array Era}

\author[]   
{S. A. Mao$^1$}

\affiliation{$^1$Max Planck Institute for Radio Astronomy, Auf dem H\"ugel 69, \\ D-53121 Bonn, Germany 
\\ email: {\tt mao@mpifr-bonn.mpg.de} \\[\affilskip]}

\pubyear{2018}
\jname{Astronomy in Focus, Volume 1} 
\editors{Piero Benvenuti, ed.}
\begin{document}

\maketitle

\begin{abstract}
The unprecedented sensitivity, angular resolution and broad bandwidth coverage of Square Kilometre Array (SKA) radio polarimetric observations will allow us to address many long-standing mysteries in cosmic magnetism science. I will highlight the unique capabilities of the SKA to map the warm hot intergalactic medium, reveal detailed 3-dimensional structures of magnetic fields in local galaxies and trace the redshift evolution of galactic magnetic fields. 
\end{abstract}
\firstsection 
\section{Introduction}

The Square Kilometre Array (SKA) will be the most powerful radio telescope in the world and it is currently in pre-construction phase. It will be hosted at two separate sites: SKA1-MID (350 MHz$-$24 GHz) with 133 15-m SKA dishes and 64 13.5-m MeerKAT dishes over a maximum baseline of 150 km will be located in the Karoo site in South Africa, while SKA1-LOW (50$-$350 MHz) with 130,000 antennas over a maximum baseline of 65 km will be located in the Boolardy site in Western Australia. The full SKA1-MID will have $\sim$ 5 times better sensitivity and 4 times higher angular resolution than the Karl G. Jansky Very Large Array, while the full SKA1-LOW will be a factor of $\sim$ 8 more sensitive than the Low Frequency Array. Science commissioning of SKA1 will commence in 2022 and full operation will begin in 2025$^{\footnotemark[1]}$\footnotetext[1]{For the anticipated SKA1 science performance, see document number: SKA-TEL-SKO-0000818 (Braun, Bonaldi, Bourke, Keane \& Wagg 2017). For the current timeline, see document number: SKA-TEL-SKO-00000822 (SKAO Science \& Op Teams 2017). Both documents are available on https://astronomers.skatelescope.org/documents/.}. The origin and evolution of cosmic magnetism is one of the five original SKA Key Science Projects (\cite[Gaensler, Beck \& Feretti 2004]{gbf2004}). In the updated SKA science book {\it Advancing Astrophysics with the SKA}, the magnetism community continues to have a strong presence, contributing a total of 19 chapters. This proceedings will highlight selected topics covered in these chapters, as well as latest developments in the field which were not included in the science book.

Radio polarization observations encode rich information on particle densities and magnetic fields in the Universe on different scales: from Mpc down to sub-pc scales. Besides mapping the polarized synchrotron emission from the astrophysical object of interest, a key measurement of cosmic magnetism is the Faraday rotation towards polarized background extragalactic radio sources. These measurements, forming a so-called rotation measure (RM) grid, enable us to directly probe the magnetic field strength and direction, as well as the gas density in the foreground intervening medium.

Our knowledge of the rotation measure sky has improved significantly in the past 20 years. In the early 2000s, only $\sim$10$^3$ extragalactic sources have Faraday rotation measurements (\cite[Johnston-Hollitt 2003]{mjh2003}). At present, the NRAO VLA Sky Survey rotation measure catalog of \cite[Taylor \etal\ (2009)]{taylor2009} (1 source deg$^{-2}$ at DEC$>$$-$40$^\circ$) along with the S-PASS/ATCA catalog of \cite[Schnitzeler \etal\ (2018)]{s2018} (0.2 source deg$^{-2}$ at DEC$<$0$^\circ$) provide us with $\sim$40,000 extragalactic RMs across the entire sky, enabling magnetic field measurements in a range of different foreground astrophysical objects. In addition to the ever increasing density of the all-sky RM grid, the advent of broadband polarimetry --- a more than 10-fold increase in the instantaneous observing bandwidth in frequency --- have brought revolution to our field in the recent years as well. Precise and n$\pi$-ambiguity free Faraday rotation (\cite[Ma \etal\ 2018]{ma2018}) along with other properties of the magnetized gas can be derived using newly-developed broadband polarization analysis tools.

An all-sky full Stokes survey at 2$"$ resolution with SKA1-MID (Band 2) down to 4 $\mu$Jy beam$^{-1}$ will provide $\sim$ 7 to 14 million extragalactic radio sources with Faraday rotations (\cite[Johnston-Hollitt \etal\ 2015]{mjh2015}). This dense RM grid will facilitate the characterization of astrophysical magnetic fields, gas densities and turbulence in unprecedented details.

\section{Extragalactic Magnetism Science with the SKA}

\subsection{Revealing the elusive missing baryons and the magnetic fields in the cosmic web}
Only about 50\% of the expected baryons in the Universe can be accounted for, while the rest -- missing baryons -- are  thought to reside in the warm-hot intergalactic medium (WHIM) in the form of shock-heated gas at 10$^5$$-$10$^7$K. The WHIM can be traced by absorption lines in X-ray and UV spectra, but extremely long integration time is required to produce high significance detection along a single sight line (\cite[Nicastro \etal\ 2018]{n2018}). Radio observations offer a highly complementary approach to characterize the particle density and magnetic fields in the cosmic web. Shocks produced by the accretion of baryonic matter as large-scale structures of the Universe form are sufficient to accelerate particles to relativistic energies, illuminating the cosmic web in synchrotron emission in the presence of intergalactic magnetic fields. The WHIM should also produce imprints on the RM of background sources if magnetic fields permeate the cosmic web.  

Currently, limits on the surface brightness and the magnetic fields of the cosmic web can be placed by cross-correlating tracers of large-scale structures and diffuse synchrotron emission (e.g., \cite[Brown \etal\ 2017]{brown2017}, \cite[Vernstrom \etal\ 2017]{vernstrom2017}). Intergalactic magnetic field strength can also be estimated by determining whether a difference in the number of large scale structure filaments intercepting the line-of-sight corresponds to a difference in Faraday rotation between two nearby sight lines (\cite[O'Sullivan \etal\ 2018]{os2018}). A potentially new population of radio sources in environments connecting galaxy clusters was discovered recently (\cite[Vacca \etal\ 2018]{vacca2018}). These sources have properties similar to those expected from the brightest patches of the diffuse emission associated with the WHIM. 

With the advent of the SKA comes new methods to reveal the cosmic web. Direct imaging of the brightest filaments of the cosmic web will be possible with deep ($>$1000 hours) SKA1-LOW observations (\cite[Vazza \etal\ 2015]{vazza2015}), provided that confusion from Galactic foreground synchrotron emission and extragalactic point source emission are minimized and accurately removed. A dense ($>$10$^3$ sources deg$^{-2}$) and precise rotation measure grid (error$\sim$1 rad m$^{-2}$) towards filaments of the cosmic web and its structure function can probe the turbulent scale, gas densities and magnetic fields in the intergalactic medium (\cite[Akahori \& Ryu 2011]{akahori2011}, \cite[Taylor \etal\ 2015]{taylor2015}). Performing a joint Faraday rotation and dispersion measure analysis of the $\sim$10$^{4}$ localized fast radio bursts anticipated by the SKA (\cite[Macquart \etal\ 2015]{macquart2015}) can put tight limits on properties of the magnetized intergalactic medium if the {\it in situ} and host galaxy contributions can be reliably subtracted (\cite[Akahori \etal\ 2016]{akahori2016}, Johnston \etal\ this volume). The all-sky RM grid with redshift information will enable the employment of new Bayesian algorithms (\cite[Vacca \etal\ 2016]{vacca2016}) to statistically isolate Faraday rotation produced by the cosmic web. These different approaches will yield densities and magnetic fields of the WHIM, addressing the missing baryon problem and ultimately distinguishing between different magnetogenesis scenarios.

\subsection{Mapping the 3-dimensional magnetic fields in nearby galaxies}

The leading theories of the amplification of magnetic fields in galaxies are the fluctuation dynamo and the large-scale $\alpha$-$\Omega$ dynamo, but details of these processes remain poorly constrained observationally. Rigorous comparisons between theoretical predictions and observations of 3-dimensional magnetic fields in nearby galaxies are necessary to fully understand these dynamos. Magnetic fields in approximately 100 galaxies have been studied by measuring their diffuse polarized synchrotron emission at limited angular resolution with mostly narrowband data (\cite[Beck \& Wielebinski 2013]{bw2013}). Only 3 nearby large-angular extent galaxies, the Large and the Small Magellanic Clouds and M31, have had their magnetic fields probed via the RM-grid approach (\cite[Gaensler \etal\ 2005]{gaensler2005}, \cite[Mao \etal\ 2008]{mao2018}, \cite[Han \etal\ 1998]{han1998}). These studies have established some basic properties of galactic magnetic fields, such as the typical field strength and the dominant disk field symmetry, but a 3-dimensional picture of galactic magnetic fields is still lacking.
Broadband radio polarimetry in combination with models of the magnetized interstellar medium (ISM) can be used to conduct tomography to characterize both large and small-scale galactic magnetic fields as a function of the line-of-sight depth, thus yielding 3-D pictures of galactic magnetic fields (\cite[Fletcher \etal\ 2011]{fletcher2011}, Kierdorf \etal\ this volume). 

While until now only a few galaxies have broadband polarization data that allow for tomography studies, the SKA will revolutionize this area. With the SKA, sensitive diffuse polarized synchrotron emission can be measured with excellent $\lambda^2$ coverage, specifically with SKA1-MID band 2, 4 and possibly 3 (\cite[Beck \etal\ 2015]{beck2015}, \cite[Heald \etal\ 2015]{heald2015}), which is ideal for magnetic field tomography. The SKA will have enough surface brightness sensitivity to map the polarized emission at extremely high angular resolution (1 kpc at $z$$\sim$0.04). Moreover, at least 200 nearby galaxies will have enough polarized background sources for us to conduct RM-grid experiments on. A joint analysis of the broadband diffuse polarized emission and the RM grid of nearby galaxies will provide a complete 3-D view of their magnetic fields. The strength and structure of the disk and halo magnetic fields and their radial and vertical dependencies can be determined. The nature (isotropic vs. anisotropic) and the power spectrum of random magnetic fields can be derived. With much improved angular resolution, an extensive search for extragalactic large-scale magnetic field reversals will also be feasible. A clear link between galaxy properties and their magnetic fields will emerge from these SKA data and will allow us to constrain the field generation processes.

\subsection{Tracing the redshift evolution of galactic magnetic fields}
Since galactic magnetic fields play important roles in processes that are closely linked to galaxy evolution, it is crucial to understand how galaxies and their magnetic fields have co-evolved over cosmic time. Directly tracing the redshift evolution of galactic magnetic fields is a challenging task: polarized synchrotron emission from cosmologically distant galaxies is faint and Faraday rotation produced by these distant galaxies when seen against background polarized sources is subjected to redshift dilution and is difficult to isolate from other sources of Faraday rotation along the line of sight. As a result, measurements of magnetic fields in galaxies beyond the local Universe are scarce. Recently, \cite[Mao \etal\ (2017)]{mao2017} have demonstrated that strong gravitational lensing of polarized background quasars by galaxies offers a clean and effective probe of the $in~situ$ magnetic fields in individual cosmologically distant galaxies. Using differential polarization properties (Faraday rotation and fractional polarization) derived from broadband observations of a lensing system at $z$=0.44, the authors have derived both the magnetic field strength and geometry in the lensing galaxy as seen 4.6 billion years ago, making it the most distant galaxy with such a measurement. 

At present, the number of systems for which this technique can be applied to is limited (Mao \etal\ in prep). With an expected discovery of  $>$10$^4$ new radio-bright gravitational lenses (a factor of $>$ 100 more than the currently known systems, \cite[McKean \etal\ 2015]{mckean2015}), the SKA will provide significantly more lensing systems that are well-suited for magnetism studies, extending measurements of magnetic fields in cosmologically distant galaxies to a much wider range in redshift and in mass. Along with other tracers of galactic magnetic fields at high redshifts (e.g., \cite[Basu \etal\ 2018]{basu2018}), the SKA will enable one to firmly establish the observational trend of galactic magnetic fields as a function of cosmic time.

\section{Summary}
The SKA will transform our understanding of the origin and evolution of cosmic magnetic fields. The dense, broadband all-sky RM-grid together with additional pointed observations of selected targets will greatly advance our knowledge on the magnetized WHIM, 3-D magnetic fields in galaxies and their redshift evolution.


\begin{thebibliography}{}

\bibitem[Akahori \& Ryu 2011]{akahori2011}
{Akahori, T. \& Ryu, D.} 2011,
\textit{ApJ}, 738, 134

\bibitem[Akahori \etal\ 2016]{akahori2016}
{Akahori, T., Ryu, D., \& Gaensler, B. M.} 2016,
\textit{ApJ}, 824, 105

\bibitem[Basu \etal\ (2018)]{basu2018}
{Basu, A., Mao, S. A., Fletcher, A., et al.} 2018,
\textit{MNRAS}, 477, 2528

\bibitem[Beck \& Wielebinski 2013]{beck2013}
{Beck, R., \& Wielebinski, R.} 2013,
{in Planets, Stars and Stellar Systems, Vol. 5, ed. T. D. Oswlat \& G. Gilmore (Dordrecht: Springer)}, 641, updated in 2018 (arXiV: 1302.5663)

\bibitem[Beck\etal\ (2015)]{beck2015}
{Beck, R., Bomans, D., Colafrancesco, S.,  et al.} 2015,
\textit{PoS, AASKA14}, 94

\bibitem[Brown \etal\ 2017]{brown2017}
{Brown, S., Vernstrom, T., Carretti, E., et al.} 2017,
\textit{MNRAS}, 458, 4246

\bibitem[Fletcher \etal\ 2011]{fletcher2011}
{Fletcher, A., Beck, R., Shukurov, A., Berkhuijsen, E. M., \& Horellou, C.} 2011,
\textit{MNRAS}, 412, 2396

\bibitem[Gaensler, Beck \& Feretti (2004)]{gbf2004}
{Gaensler, B. M., Beck, R., \& Feretti, L.} 2004,
\textit{New Astron. Revs}, 48, 1003

\bibitem[Gaensler \etal\ 2005]{gaensler2005}
{Gaensler, B. M., Haverkorn, M., Staveley-Smith, L., et al.} 2005,
\textit{Science}, 307, 1610

\bibitem[Han \etal\ 1998]{han1998}
{Han, J. L., Beck, R., \& Berkhuijsen, E. M. } 1998,
\textit{A\&A}, 335, 1117

\bibitem[Heald \etal\ (2015)]{heald2015}
{Heald, G., Beck, R., de Blok, W. J. G., et al.} 2015,
\textit{PoS, AASKA14}, 106

\bibitem[Johnston-Hollitt (2003)]{mjh2003}
{Johnston-Hollitt, M.} 2003, 
\textit{Phd Thesis}, University of Adelaide

\bibitem[Johnston-Hollitt \etal\ (2015)]{mjh2015}
{Johnston-Hollitt, M., Govoni, F., Beck, R., et al.} 2015,
\textit{PoS, AASKA14}, 92

\bibitem[Macquart  \etal\ 2015]{macquart2015}
{Macquart, J. P., Keane, E., Grainge, K., et al.} 2015,
\textit{PoS AASKA14}, 55

\bibitem[Ma \etal\ 2018]{ma2018}
{Ma, Y. K., Mao, S. A., Stil, J., et al.} 2018,
\textit{MNRAS}, submitted

\bibitem[Mao \etal\ 2008]{mao2008}
{Mao, S. A., Gaensler, B. M., Stanimirovi\'c, S., et al.} 2008,
\textit{ApJ}, 688, 1029 

\bibitem[Mao \etal\ (2017)]{mao2017}
{Mao, S. A., Carilli, C., Gaensler, B. M., et al.} 2017,
\textit{Nature Astronomy}, 1, 621

\bibitem[McKean \etal\ (2015)]{mao2015}
{McKean, J., Jackson, N., Vegetti, S., et al.} 2015,
\textit{PoS, AASKA14}, 92

\bibitem[Nicastro \etal\ 2018]{n2018}
{Nicastro, F., Kaastra, J., Krongold, Y., et al.} 2018,
\textit{Nature}, 558, 406

\bibitem[O'Sullivan \etal\ 2018]{os2018}
{O'Sullivan, S. P., Machalski, J., Van Eck, C. L., et al.} 2018,
\textit{MNRAS}, submitted

\bibitem[Schnitzeler \etal\ (2018)]{s2018}
{Schnitzeler, D. H. F. M., Carretti, E., Wieringa, M. H., et al.} 2018, 
\textit{MNRAS}, submitted

\bibitem[Taylor \etal\ (2009)]{taylor2009}
{Taylor, A. R., Stil, J. M., \& Sunstrum, C.} 2009, 
\textit{ApJ}, 702, 1230

\bibitem[Taylor \etal\ 2015]{taylor2015}
{Taylor, A. R., Agudo, I., Akahori, T., et al.} 2015,
\textit{PoS AASKA14}, 113

\bibitem[Vacca \etal\ 2016]{vacca2016}
{Vacca, V., Oppermann, N., Ensslin, T., et al.} 2016,
\textit{A\&A}, 519, A13


\bibitem[Vacca \etal\ 2018]{vacca2018}
{Vacca, V., Murgia, M., Govoni, F., et al.} 2018,
\textit{MNRAS}, 479, 776

\bibitem[Vazza \etal\ 2015]{vazza2015}
{Vazza, F., Ferrari, C., Bonafede, A., et al.} 2015,
\textit{PoS AASKA14}, 97

\bibitem[Vernstrom \etal\ 2017]{vernstrom2017}
{Vernstrom, T., Gaensler, B. M., Brown, S., Lenc, E., \& Norris, R. P.} 2017,
\textit{MNRAS}, 467, 4914

 
\end{thebibliography}
\end{document}